\documentclass[12pt]{article}
\usepackage{fullpage,setspace}
\usepackage{amssymb, amsthm, amsmath}
\usepackage{graphicx}
\usepackage[authoryear]{natbib}
\usepackage{bm}
\usepackage{times}
\usepackage{setspace}
\usepackage{wrapfig}
\usepackage{epstopdf}
\usepackage{subcaption}
\usepackage{caption}
\usepackage{booktabs}
\usepackage{multirow}
\usepackage{verbatim}
\usepackage{url}
\usepackage{hyperref}

\title{Oral exams in introductory statistics class with non-native English speakers}
\author{Eric Yanchenko\\ 
Akita International University\\ Akita, Japan}
\date{\today}

\begin{document}

\maketitle

%\doublespacing

\begin{abstract}
    \noindent
    Oral exams are a powerful tool to assess student's learning. This is particularly important in introductory statistics classes where students struggle to grasp various topics like the interpretation of probability, $p$-values and more. The challenge of acquiring conceptual understanding is only heightened when students are learning in a second language. In this paper, I share my experience administering oral exams to an introductory statistics class of non-native English speakers at a Japanese university. I explain the context of the university and course, before detailing the exam. Of particular interest is the relationship between exam performance and English proficiency. The results showed little relationship between the two, meaning the exam seemed to truly test student's statistical knowledge rather than their English ability. I close with encouragements and recommendations for practitioners hoping to implement similar oral exams, focusing on the unique difficulties faced by students not learning in their mother tongue. 
\end{abstract}

\section{Introduction}

Oral exams are a unique tool to assess student's understanding of course material. This is especially relevant in introductory statistics courses where many topics are conceptually difficult, e.g., interpretation of probability, $p$-values, hypothesis testing, etc. These challenges are only heightened for students learning in a second language. Specifically, in this work we are interested in an introductory statistics course of predominantly Japanese students learning in English. In this context, oral exams provide many benefits for assessing understanding. First, they give instructors a clearer picture of students' conceptual mastery \citep{theobold2021oral, sabin2021oral}. As discussed in the Guidelines for Assessment and Instruction in Statistics Education (GAISE), conceptual understanding is a fundamental goal for statistics educators \citep{carver2016guidelines}. Specifically, this means instructors should ``view the primary goal as to discover and apply concepts'' and ``focus on students’ understanding of key concepts, illustrated by a few techniques, rather than covering a multitude of techniques with minimal focus on underlying ideas'' \citep{carver2016guidelines}. Unfortunately, far too many introductory statistics courses focus on formulas and memorization, with little emphasis on conceptual understanding. This is even more so the case in the Japanese context, where memorization is a major pillar of primary through higher education \citep[e.g.,][]{nemoto1999japanese}. In particular, \cite{dunn2015critical} highlights a lack of critical thinking skills in Japanese college freshmen, largely stemming from a secondary education pedagogical framework which stresses memorization. This poses a major challenge if the goal of an introductory statistics course should be to ensure students are ``able to make decisions about the most appropriate ways to visualize, explore, and, ultimately, analyze a set of data'' \citep{carver2016guidelines}. 

Unlike memorizing formulations and definitions, oral exams require students to clearly communicate statistical concepts, a necessary component to demonstrate mastery \citep{garfield2008assessment}. They also allow the instructor to probe answers to assess comprehension, which in turn, forces students to build a deeper understanding \citep{wiggins2005understanding}. The benefits work in the opposite direction as well; this testing format means students can ask for clarification when they do not understand a question \citep{ohmann2019assessment}, a frequent occurrence for students learning in a second language.

The importance of assessing mastery has only increased with the current proliferation of large language models (LLMs), e.g., ChatGPT, Gemini, etc. It is now easier than ever for students to circumvent the learning process by simply passing off their assignments to LLMs \citep{paustian2024students, al2025enhancing}. This problem is pervasive for students learning in a second language where the stress of deadlines and writing in another language may push them to use an LLM, or assignments may be completed in Japanese only to have an application translate it to English. Ideally, with the prospect of oral exams, students should be motivated to truly learn the material \citep{delson2022can} instead of inappropriately relying on LLMs. Moreover, not only do oral exams necessitate deep, conceptual understanding of statistical concepts, but they also allow students to build their oral communication skills \citep{hazen2020use, sabin2021oral}.

Oral exams can also help students prepare for their future careers. In my context, the vast majority of students will likely never take another statistics course. They will, however, encounter statistical ideas in their future careers, whether it be business, consulting, manufacturing or something else. Indeed, a student is unlikely to be asked to solve a series of problems with pencil-and-paper in their future job, but their boss may ask them to clearly explain some statistical idea. Moreover, interviews are a key component of job hunting everywhere, but particularly in Japan. In many respects, oral exams simulate the stressors of an interview setting, requiring quick thinking, clear communication and confidence.

In this paper, I share my experiences administering oral exams (in English) to an introductory statistics class at a Japanese university with non-native English speakers. After detailing the university environment, I discuss the course layout and previous exam implementations. Following this, I detail the oral exam and discuss the results. Finally, I share concluding thoughts and give advice for those hoping to administer oral exams in a similar context.

In many ways, this paper was inspired by \cite{theobold2021oral}, but there are several key differences. First, this is an introductory class while \cite{theobold2021oral} focuses on an intermediate regression course. More distinctively, however, this paper deals with giving exams to students learning in their second language. Indeed, one of the major questions I had was: will the students with the strongest English skills perform the best on the exams? If so, then perhaps the exam is unfairly testing English proficiency instead of statistical ability. In general, I found little correlation between exam scores and English speaking skills. Lastly, I trace the unique challenges and opportunities of oral exams in a non-native speaking setting.

\section{Setting}

\subsection{University}
Oral exams were given to an introductory statistics class of nine students at Akita International University (AIU) in Akita, Japan during the spring 2025 semester. AIU is well-known in Japan for its unique learning environment; while the vast majority of degree-seeking students are Japanese, classes are taught entirely in English. Students are admitted with fairly strong English skills and then spend their first year honing these skills to an academic level. Specifically, students must complete the English for Academic Purposes (EAP) program to improve their reading, writing, speaking and listening skills. These courses are designed to help students improve their English proficiency such that they are prepared to learn (other subject matters) in English. Furthermore, AIU is modeled after an American liberal arts college. While many Japanese universities have become highly specialized or technical, AIU seeks to form well-rounded, globally-minded individuals.

There are key implications from these features on the statistics class. First, for every student in my class, English was not their first language. Of the nine students in the class, I judged two of them to posses noticeably better English skills than the typical AIU student, likely because these two students had spent significant time living overseas. For the remaining seven students, however, their English skills were not on par with a native speaker. I was eager to see whether there was a strong correlation between English proficiency and test performance. Second, AIU is one of the few liberal arts college in Japan. While the classical liberal arts include a heavy-dose of mathematics and sciences, there is a perception in Japan (and likely more broadly) that liberal arts means you are interested in the humanities as opposed to math and science. Thus, AIU primarily attracts students who are not keen on studying in the STEM fields. Finally, only one student had taken a college-level statistics course before.

\subsection{Course details}

For a class to use oral assessments, it is important to prepare students throughout the entire course; it is not sufficient to focus on formulas and definitions and then expect students to excel on conceptually-based oral exam questions \citep{theobold2021oral}. Thus, I prepared students for the exams through lectures, homework assignments and the use of \texttt{R} software.

Lectures were primarily focused on conceptual understanding as opposed to calculations and formulas. I stressed ideas such as: frequentist vs.~Bayesian interpretation of probability, populations vs.~samples \citep{allison2025sample}, appropriateness of tests in different scenarios and more. Students rarely had to do calculations by-hand unless it aided in a conceptual understanding of the method. Additionally, I explicitly noted when a certain topic or question would likely appear on the oral exam. 

Similarly, homework problems assessed understanding and concepts over calculation. For example, students would be given an experiment and asked to describe the population and sample. Or, students may be asked for the assumptions of a model and then encouraged to think about whether they are met. While there were not explicit oral components to the homeworks, the types of questions were similar to those which appeared on the exams. Students also had an opportunity to practice answering oral questions with each other before the exam.

Lastly, ideas were illustrated using \texttt{R} for almost every topic. The course topics generally followed {\it Learning Statistics with R} \citep{navarro2013learning}, a textbook that integrates coding with concepts and is aimed at non-statistics majors. Beyond hoping that students would gain some programming skills, the main motivation for incorporating \texttt{R} was to aid in the understanding of certain concepts. Sampling distributions, the Law of Large Numbers and the Central Limit Theorem are just some of the concepts that introductory students typically struggle with. By using \texttt{R} to ``see'' these concepts play out, I hoped that students would gain a deeper appreciation. That being said, the midterm exam did not consist of any questions about \texttt{R}.

\subsection{Previous implementations}
This was my third time conducting oral exams in this course as I had administered them during the previous two semesters. These experiences paved the way for the current implementation. In the first iteration in spring 2024, students were given an oral midterm and final exam. The midterm was ten minutes and contained two definition questions and two conceptual questions. The final exam was thirty minutes and included three questions analyzing the results of a model or test. I found the final to be extremely long, and even so, many students could not answer all of the questions in the allotted time. Thus, in the fall 2024 semester, the final exam was converted to a written exam, but the midterm format was kept roughly the same. During this second implementation, I realized that four questions was too much for only ten minutes. In particular, there was not enough time to probe student's answer as I needed to rush to the next question. Additionally, test performance was generally quite poor. Finally, there were multiple versions of the exam in these previous implementations, and I did not record them. Based on these experiences, the test was modified to its current iteration, described below.

\section{Oral midterm exam}
\subsection{Format}
The midterm exam was given approximately half-way through the semester. By this point, I had taught summary statistics, probability and some basics of sampling distributions and hypothesis testing. The midterm was worth 25\% of the final grade. Each student was given fifteen minutes to answer three questions. I read the questions to the students and at no point did they have an opportunity to read the questions themselves. The week before, students signed up for their exam time on an online spreadsheet, and students came to my office individually to take the exam. The testing process took roughly the whole day with breaks in-between students.

Before the exam, I asked students to complete a brief survey on their English abilities, asking: ``How would report your own English abilities (overall)?'' and, ``How would report your own English abilities? (Speaking)?'' Response options were: very weak, weak, intermediate, proficient and fluent. As a more objective measure, students also shared their standardized test scores. All but two students reported a TOEFL iTP score, a popular test of English proficiency for universities, so this will be the main metric in this analysis. Finally, I subjectively assessed each student's English ability and deemed two students to have above-average English speaking abilities compared to the typical student at the university. 

During the exam, I graded and took notes while also recording the exam with a smartphone application. The purpose of the recording was to ensure fair grading across students, e.g., the same answers by different students received the same grade, and as an insurance policy in case students disputed their grade and/or answers. Indeed, in a previous implementation when I did not record the exams, a student protested part of their grade but since there was no recording, it was impossible to adjust. After the exam, I also gave some brief feedback to each student on their performance. This study received Institutional Review Board (IRB) approval from my university. Lastly, the final exam in the course was a more traditional written exam.

\subsection{Questions and scoring}\label{sec:scoring}
Approximately one week before the exam, students were told some topics that might appear on the exam. All exam questions came from this list and each student was given the same exam. Each question was scored out of 4 points (with a granularity of 0.5 points). After scoring, I compared the results with the mastery rubric from \cite{theobold2021oral} (included in the Appendix) and the scores generally tended to agree. As the only instructor for the course, achieving consistent and reliable grading was difficult, but recording and re-listening to the exams helped mitigate these challenges. The exam questions are below with italicized writing briefly explaining the grading.

\begin{enumerate}
    \item Explain to someone who has never taken a statistics course what a $p$-value is. 
    
    {\it (1 point for correct definition including assumption that null is true and probability of being as extreme or more than the observed value. Also should explain the idea behind the null hypothesis.)}
    \begin{itemize}
        \item Why do we need to assume the null hypothesis is true to calculate the p-value? 
        
        {\it (1 point for discussing how the population parameter is unknown.)}
        
        \item  What does ``extreme'' mean?
        
        {\it (1 point for explaining that the expected and surprising result depends on the null and alternative hypothesis.)}
        
        \item What does a small / large p-value suggest? 
        
        {\it (1 point for saying a small $p$-value gives evidence against the null and vice-versa for a large $p$-value.)}
    \end{itemize}

    \item After an earthquake, the government reports that there is a 90\% chance of a tsunami. However, no tsunami comes so your friend thinks that the government's probability was wrong. Explain how a Bayesian and frequentist would respond to the friend.
    \begin{itemize}
        \item Bayesian interpretation

        {\it (1.5 points for correct explanation including subjective belief and updating probability after observing data.)}
        
        \item Frequentist interpretation

        {\it (1.5 points for correct explanation including long-run frequency and this situation is the 10\% chance of a tsunami not coming.)}
        
        \item Which makes more sense if we are interested in the prediction for this exact day? Why?

        {\it (1 point for discussing how Bayesian is preferable for single events.)}
    \end{itemize}

    \item Explain the idea of the Central Limit Theorem.

    {\it (1 point for correct explanation, including sample mean distribution, normal distribution and large sample size.) }
    \begin{itemize}
        \item What does the CLT tell us about the mean of the sample mean distribution? What about the variance?

        {\it (1 point for mean is population mean and variance is population variance divided by sample size.)}
        
        \item What is the difference between the sample and the population mean?

        {\it (1 point for correct distinction including population mean is unknown and sample mean is known.) }
        
        \item Does data need to be normal?

        {\it (1 point if knows that the data does not need to be normal.)}
    \end{itemize}
\end{enumerate}

\subsection{Results}\label{sec:results}
\begin{figure}
    \centering
    \includegraphics[width=0.85\linewidth]{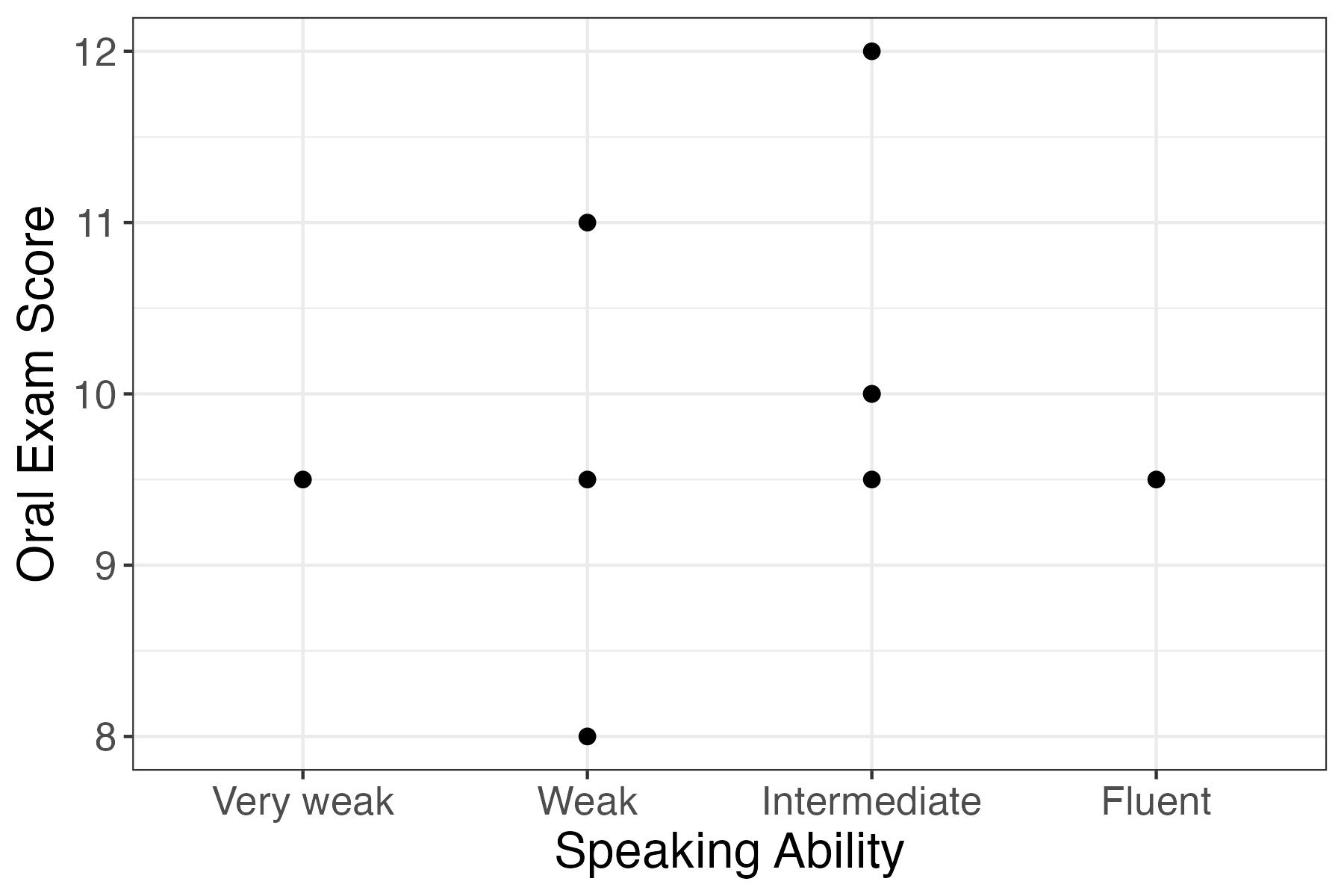}
    \caption{Scatter plot of oral midterm exam performance (out of twelve points) against self-assessed speaking ability.}
    \label{fig:hist}
\end{figure}

The oral midterm exam performance for each student is reported in Figure \ref{fig:hist} along with their self-assessed speaking ability. The average score (out of twelve points) was $9.9$ with a standard deviation of $1.1$. The median and mode were both $9.5$. Separated by individual question, the average (out of four) and standard deviation for the $p$-value, probability interpretation and Central Limit Theorem questions were $2.9$ ($0.7$), $3.8$ ($0.3$) and $3.2$ ($0.6$), respectively. Additionally, the correlation between the oral midterm and (written) final exam scores was $0.80$.

\begin{figure}
    \centering
    \includegraphics[width=0.85\linewidth]{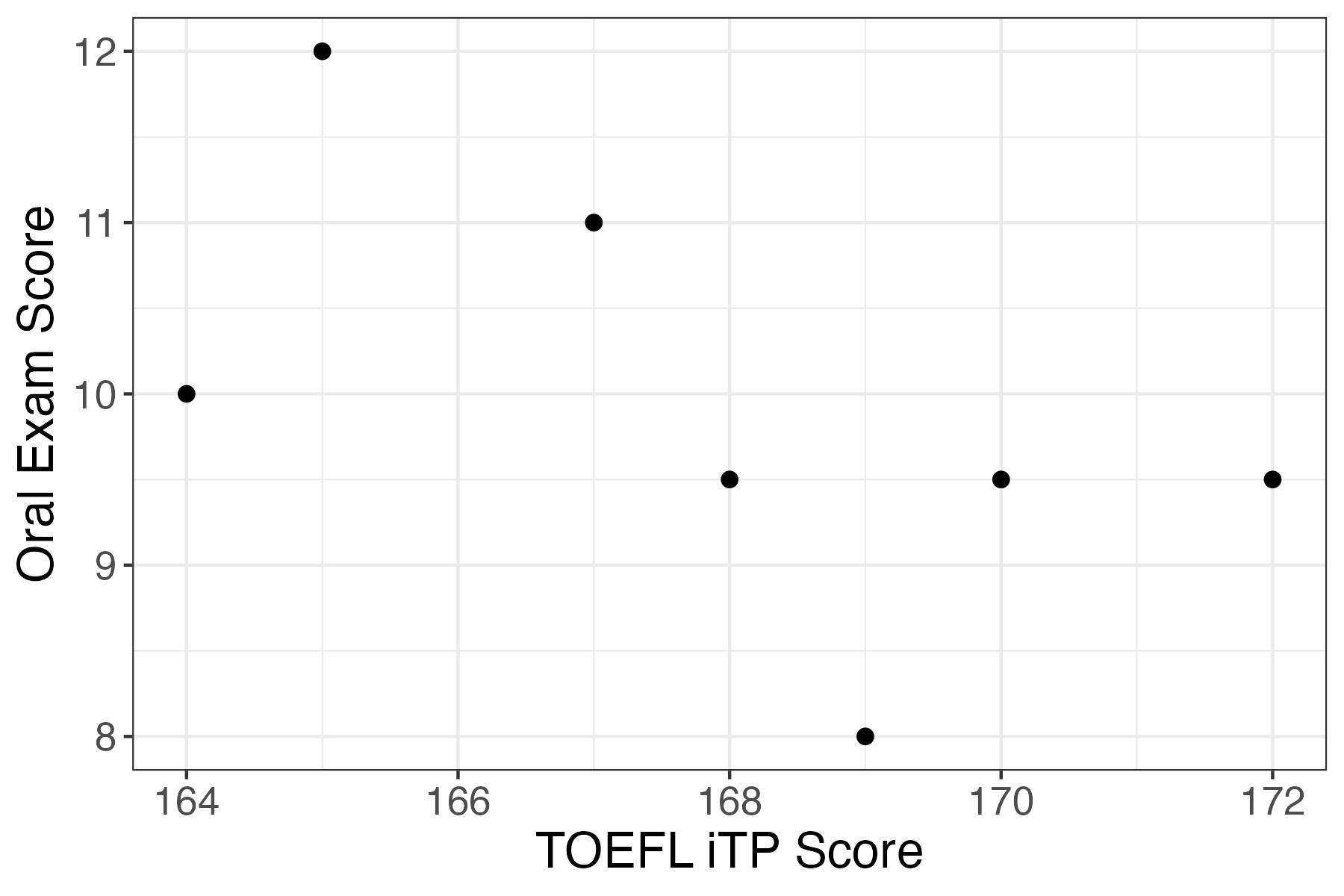}
    \caption{Scatter plot of oral midterm exam performance (out of twelve points) against TOEFL iTP score (out of $203$ points).}
    \label{fig:scatter}
\end{figure}

In Figure \ref{fig:scatter}, the student's oral exam score and TOEFL iTP score (out of $203$ points) are plotted for the seven students who reported this value. Based on their self-assessed speaking ability, four students said they had ``weak'' or ``very weak'' speaking skills, while the remaining five choose ``intermediate'' or ``fluent.'' The students in the former group had a mean score of $9.5$, while the mean of the latter group was $10.2$. On the final exam, these means were $78$ and $82$ (out of $100$), respectively. Finally, three of the top four scores were from students with typical English ability compared to others at the university, based on my subjective assessment of the student's language ability.

\subsection{Example responses}
The following is a fairly representative example of a student's explanation of the $p$-value which included all of the main points: ``$p$-value is the probability that I can get the result which is as extreme or more extreme than I observed, if I assume that the null hypothesis is true. And the null hypothesis\dots is the statement which is `there is no effect' or `there is [sic] no differences.''' 

This next exchange occurred when asking what ``extreme'' means in the case of $p$-values. Unfortunately, this student omitted discussion of the null and alternative hypotheses, and mistook the null distribution with the empirical histogram of the sample:\\

\noindent
{\it Student:} Extreme means like if the data is more unlikely observed [sic]... it's far from the point that is more likely to be observed.\\

\noindent
{\it Professor:} How do we know what is likely to be observed?\\

\noindent
{\it Student:} If we see the distribution, if we have the observed data here [draws with finger on the table], then this distribution around here [referring to mode of hand-drawn distribution] is more likely to be observed.\\

\noindent
{\it Professor:} How do I know where to draw the distribution?\\

\noindent
{\it Student:} Calculate from the sample.\\

For the probability interpretation question, the following student demonstrated good understanding since he/she could explain why each interpretation might be preferable: ``For us individuals... the Bayesian theory should be more reasonable for us because the... earthquake which might result in [a] tsunami is not [a] frequent event for individuals. But for the government or scientists, the probability of tsunami coming should be calculate [sic] by lots of similar conditions or similar data so the frequentist theory should be more reasonable.'' This answer emphasizes that, for individuals, experiencing an earthquake or tsunami is likely (or hopefully) a rare event which cannot be repeated, but for the government which may be tracking seismological activity over decades or centuries, it is more reasonable to think about this as a repeatable event.

Finally, the following response to the Central Limit Theorem question is an example of probing a student's vague answer. The student said: the main idea of the Central Limit Theorem is that ``the sampling distribution of the sample mean approaches a normal distribution if we can get a large number of the sample mean.'' At this point, it was not clear whether the student meant that we have a large number of observations in our sample, or if we collect a large number of samples, so I asked, ``Can you explain what you mean by `a large number of the sample mean?''' The student then responded with ``it is related to the law of large numbers. The larger the sample size the more accurate data we can get. So for Central Limit Theorem to work, we need enough samples.'' This essentially clarified that the student understood we need a larger number of observations in a single sample, though admittedly, further probing may have still been warranted.

\subsection{Student Feedback}
After the exam, students completed a brief survey. There were two question on a scale of 1-5 with 1 being completely disagree and 5 being completely agree: ``I feel like the oral exam fairly assessed my knowledge of the material'' and, ``I feel like my English ability made it difficult to share my statistical knowledge.'' There was also space for open-ended comments. 

Eight out of nine student gave a 4 or 5 that the test fairly assessed their statistical ability while seven out of nine said their English skills hindered their ability to share statistical concepts (score of 4 or 5). As for the comments, most students shared that they were nervous and found it difficult to express their ideas. One student said, ``Actually I feel like I understand most of the material taught during the classes and I can also apply them [sic], but sometimes it is harder to explain something in a way that people understand, or even for myself to understand.''

Several students said, however, that they thought the oral exam helped them to identify weak points in their understanding. For example: ``after finishing the test, I thought [the] oral test is good to make sure how I understand what I learn because it was good chance to output my knowledge'' and, ``the oral exam clarifies what I understand and what I do not understand.'' Students also liked that they could easily ask if they did not understand the question: ``one thing good [sic] about [the] oral exam is that I can ask [the] professor when I cannot fully understand the meaning or purpose of the question. Also, [the] professor ask[ed] me further explanation to confirm my understanding, and then I can describe more to tell my interpretation.'' 

The only real complaint was from a student who felt the students who took the exam later in the day had an advantage because they could ask their friends for the questions. Of course, students were not supposed to do this, but it might have inevitably occurred. Even so, the nature of the oral exam means students still must clearly explain their answers, even if they know the questions beforehand. In general, the feedback was quite positive: ``it was\dots a great experience. Thank you.''

\section{Conclusion}

\subsection{Main findings}
In this work, I describe my experience giving oral exams in an introductory statistics class of non-native English speakers at a Japanese university. The particular focus is on the effects of students learning in their second language and, more specifically, if students with stronger English skills had an advantage on the exam. There were several major takeaways from this experience.

First, the scores were generally much higher than in previous semesters. This is likely because I gave more specific information about what to expect on the exam, i.e., example questions. Additionally, I may have graded more leniently since I was calibrated to the expected responses. Next, the most difficult questions were: ``why we need to assume the null hypothesis is true to calculate the $p$-value?'' and, ``what is the mean and variance of the normal distribution in the Central Limit Theorem?'' Since I told students that there would not be calculations on the exam, they may have have supposed that these types of questions were also off-limits, leading to the lower performance. 

One of the main advantages of oral exams is the ability to probe unclear answers to truly assess student's understanding. An example of this was detailed in the previous sub-section with the Central Limit Theorem question. In general, when I probed, I was looking for students to use the precise language we discussed in-class, but as the probability response showed (contrasting the individual and government's perspective), sometimes a student's organic response demonstrated a deeper understanding than my expectation. An unforeseen benefit of the exams was that the ability to probe unclear responses also works in the opposite direction; students can ask for clarification when {\it they} do not understand the question, as noted in the student feedback. Indeed, I noticed that in many written questions, both in this course and others, students often misunderstand the question, so this was an unexpected bonus of using oral exams, and aligned with \cite{ohmann2019assessment} who also found students benefiting from the conversation style of the exam.

As for the main research question, the results of Section \ref{sec:results} show almost no relationship between English proficiency and test scores. For those who reported their TOEFL iTP score, the higher scoring students on this standardized test actually performed slightly worse on the oral exam. Also, one of the top four scores came from a student with above-average English ability compared to his/her peers. Additionally, based on student's self-perception of their English abilities, those who were more confident in their ability scored $0.7$ points ($6\%$) higher on average. A comparable difference of $4\%$ was found on the final (written) exam as well. Moreover, the high correlation between the written and oral exam grades further supports the claim that the oral exam was testing statistical ability and not English speaking proficiency.

One surprising finding was that the better English speakers tended to get more flustered when they did not completely understand a question. For example, in the probability interpretation question, a few students were particularly confused by the phrase ``what would you say to your friend?'' It is unclear if this negatively affected these students, but their reaction was notably different from the other students. I hypothesize that the weaker English students may not have fully understood the question and resorted to answering about the topic more generally, while the stronger English students wanted to precisely answer the question. In summary, these results seem to indicate that oral exams are a fair assessment tool even with students of varied English abilities. In other words, oral exams appear to truly test statistical understanding instead of merely English proficiency, and this should give instructors the confidence to experiment with them in their own courses.

These findings are an important contribution to the English-medium instruction (EMI) literature. For example, many works assert that EMI can be detrimental to student's content learning, but there has been minimal quantitative proof \citep{macaro2018systematic}. Here, we showed that there does not appear to be a large effect on student's content learning based on their English proficiency. We acknowledge, however, that the instructor was a native English speaker, so there may be added difficulties if that is not the case \citep[e.g.,][]{tange2010caught, bradford2016toward}.

\subsection{Recommendations and limitations}
Overall, I was fairly pleased with the results of the oral exams. I would give several points of consideration to instructors wishing to implement similar exams, especially in a context with non-native English speakers. First, I encourage giving similar practice problems and low-stakes practice opportunities, as well as sufficient hints about the exam content. Even with repeated statements about the exam material, many students still could not clearly articulate a deep understanding of these concepts, e.g., why we assume the null hypothesis is true. In the non-native English context where listening and speaking comprehension skills are lower, it is vital to give sufficient reminders. 

When working with non-native speakers, I also found it imperative to give an adequate amount of time for the exam. I thought that fifteen minutes would be enough based on my previous implementations, but it was still insufficient for many students. As non-native speakers tend to take more time to understand the question, including requiring it to be read slowly and multiple times, it is important to plan sufficient time. Indeed, several students were pressed for time in a way that meant I could not ask all the probing follow-up questions that I would have liked, especially on the last question. Additionally, I suggest ``priming'' students for the type of question you will ask next during the exam to aid in their comprehension. For example, saying something like ``this next question will be about Bayesian and frequentist probability interpretations'' can help students follow the ensuing question more easily without needing it to be repeated.

I also found the grading to be more difficult than I expected, particularly for partially correct answers. For example, students might have said an extreme result was more ``surprising'' but could not clearly articulate what this meant. Moreover, it was challenging to keep consistent grading expectations throughout the day when the first and last exam were close to eight hours apart. For example, I started to realize that one of the Central Limit Theorem follow-up questions was particularly confusing for students, so half-way though the day, I changed what I was looking for in this response. Since I could re-listen to the previous responses and adjust their grades accordingly, this was not a problem, but still it created some challenges for fairness. If you plan to give an oral exam, the more precise and granular the rubric you can make, the better.

One interesting future research direction would be to compare student's performance on an oral exam that included questions both in Japanese and English. This would allow for a more direct isolation of the language effects. Of course, this would require the test proctor to be fluent in both English and Japanese, and there would still be the challenge that students were learning in their second-language (English), even if the test was partially in their primary language (Japanese). Similar studies could also be carried out in other second-language learning contexts, particularly where the mother-tongue is closer to English, e.g., Spanish.

Perhaps the biggest limitation of this study was the small sample size ($n=9$), leading to low power. That being said, if we consider the data in Figure \ref{fig:scatter} and test $H_0:\rho\geq0$ vs. $H_1:\rho <0$ where $\rho$ is the Pearson correlation coefficient, then the resulting $p$-value is 0.09, providing moderate evidence against the null hypothesis that stronger English skills lead to better oral exam performance. In any case, instead of these results being interpreted as conclusive support that oral exams do not favor students with strong English-speaking abilities, they should be taken as illustrative and an encouragement for other instructors to experiment with oral exams. 

In summary, the test did not merely gauge a students' English ability, but yielded a meaningful picture of their statistical understanding. Even while working in their second-language, I was impressed by many of the students ability to express complex statistical concepts. Moreover, the exam setting yielded the added benefit of being able to immediately encourage students who did particularly well, easing much of their stress as they left my office.

\section*{Acknowledgments}
I would like to thank Sunghwan Byun, Herle McGowan and Clay Williams for their helpful comments. I also thank the two anonymous referees and editor whose comments greatly improved the quality of the manuscript.

\section*{Data Availability}
Data is private and cannot be shared with this article.

\bibliographystyle{apalike}
\bibliography{refs}

\section*{Appendix}
The following is a modified mastery rubric, adapted from \cite{theobold2021oral}. While the exam was primarily graded based on the explanation in Section \ref{sec:scoring}, this rubric was used as a {\it post hoc} method to check the scores.
\begin{itemize}
    \item 4 points (excellent) – Outstanding ability to articulate the concepts covered in the exam, with comprehensive and thoughtful understanding of the content of the assessment. Few if if any errors occur. Any errors that do occur are minor and do not cloud the understanding articulated during the exam.
    \item 3 points (good) – Ability to articulate the central concepts covered in the exam, but several errors may occur during the assessment. These errors, however, do not cloud the central concepts of the assessment.
    \item 2 points (fair) – Limited ability to articulate the concepts covered in the exam, explanations contain inconsistencies and demonstrate limited understanding of the content
    \item 1 point (poor) – Little to no articulation of concepts covered in the exam, numerous inconsistencies and errors occur and demonstrate little to no understanding of the content
    \item 0 points (no credit) - Unable to demonstrate any understanding of concepts
\end{itemize}

\end{document}